\documentclass[aps,prx,floatfix,twocolumn,superscriptaddress]{revtex4}
\usepackage{epsfig}
\usepackage{amsmath}
\usepackage{xcolor}
\usepackage{amsfonts}

\newcommand{\ignore}[1]{}

\newcommand{\mean}[1]{\left\langle #1\right\rangle} 

\begin{document}
\title{Altruism in populations at the extinction transition}
\author{Konstantin Klemm} \email{klemm@ifisc.uib-csic.es}
\affiliation{
Institute for Cross-Disciplinary Physics and Complex Systems IFISC (UIB-CSIC), 07122 Palma de Mallorca, Spain
}
\affiliation{
Bioinformatics Group, Department of Computer Science,
Universit\"{a}t Leipzig, H\"{a}rtelstr.\ 16-18, 04107 Leipzig, Germany
}
\author{Nagi Khalil} \email{nagi.khalil@urjc.es}
\affiliation{Escuela Superior de Ciencias Experimentales y Tecnología (ESCET) \& GISC, Universidad Rey Juan Carlos, Móstoles 28933, Madrid, Spain}

\date{\today}
\begin{abstract}
We study the evolution of cooperation as a birth-death process in spatially extended populations. The benefit from the altruistic behavior of a cooperator is implemented by decreasing the death rate of its direct neighbors. The cost of cooperation is the increase of a cooperator's death rate proportional to the number of its neighbors. When cooperation has higher cost than benefit, cooperators disappear. Then
the dynamics reduces to the contact process with defectors as the single particle type. Increasing the benefit-cost ratio above 1, the extinction transition of the contact process splits into a set of nonequilibrium transitions between four regimes when increasing the baseline death rate $p$ as a control parameter: (i) defection only, (ii) coexistence, (iii) cooperation only, (iv) extinction. We investigate the transitions between these regimes. As the main result, we find that full cooperation is established at the extinction transition as long as benefit is strictly larger than cost. Qualitatively identical phase diagrams are obtained for populations on square lattices and in pair approximation.  Spatial correlations with nearest neighbors only are thus sufficient for sustained cooperation.
\end{abstract}
\maketitle

\section{Introduction}

Altruism or cooperativity \cite{tr71} describe behavior that is more in favor of others than of the actor herself. Alarm calls are an example of altruistic behavior: Increasing the risk of becoming prey itself first, one individual of a group warns the others of a predator approaching \cite{ta87}. At first glance, the observation of altruism sustained over generations appears incompatible with Darwin's theory of natural selection, featuring the survival of the fittest \cite{darwin1859,spencer1864}. If non-altruists acting only to their own benefit have an advantage over altruists in terms of reproductive success, altruistic traits eventually disappear.

The question of sustained altruism and cooperativity has been addressed in the framework of evolutionary game theory, in particular by work on the Prisoner's Dilemma and Public Goods Games \cite{da80,axha81}. In these and other games, the time evolution of the system is assumed to take place as a sequence of two elementary steps: (i) the combined behavioral choices of the participants lead to an assignment of a payoff to each player which (ii) determines the choice of their future strategies or roles. In the simplest case, with two possible strategies, cooperation and defection, the dilemma arises as follows. Regardless of the other agent's move, an agent's best (highest payoff) move is always defection. On the other hand, the sum of all players payoffs is maximal when all cooperate. Therefore, natural selection always favors defection \cite{da80}, despite cooperation is the best global strategy.
 
The aforementioned social dilemma is frequently analyzed by means of the replicator equation \cite{ohno06,rocusa09,brpa18} describing the time evolution of the fraction of players holding one of the two strategies. If the fitness of an individual equals its payoff, the resulting replicator equation for the Prisoner's Dilemma has only two steady-state solutions, the only-defector and the only-cooperator solutions, the former being the only stable one. Nevertheless, the prevalence of cooperation is still possible within the context of evolutionary games, provided appropriate reciprocity mechanisms are included in the dynamics \cite{tr71,axha81,leke06,za14,wakojuta15}: Direct reciprocity, indirect reciprocity, kin selection, group selection, and network structure. If compared to the well-mixed situation, the new mechanisms include update rules that favor the interactions among cooperators.    

The network structure mechanism was one of the first reciprocity mechanisms studied in the literature. It refers to the restriction of agents interactions among neighbors. In a two-dimensional regular network, the survival of altruists was explained in terms of their ability of preventing the exploitation of defectors through the formation of clusters \cite{noma92,namaiw97,tamaiw97,iwnale98}. Further progress in the field considered births and deaths: The second step of the dynamics, the one that allows a change of the strategy, is now interpreted as a death of a player followed by a birth. The new ecologic perspective allowed to assess the importance of new relevant issues, such as the fluctuation of the population density \cite{miwi00,lilihiiwboni13,Huang:2015,mcfrhawano18}, the movement of agents \cite{bara98,yiliwalulimali14,busz05,vasiar07,lero10}, the spatial distribution of neighbors and their number \cite{ifkido04,sena09}, among others. Recent works also consider networks of interactions \cite{ohno06,asgola08,fuwanoha09,notaan10,pegoszflmo13,wakojuta15,kiyoki16}, focus on the critical properties of the system \cite{hasz05,szvusz05,vuszsz06}, include other novel dynamic rules \cite{raamnaoh10,pesz10,pisapa12,liliclgu15,pachadno15,iyki16,szpe17}, analyze the formation of patterns  \cite{noma92,szfa07,lanoha08,funoha10,waha11,yabasa18}, and evaluate the effect on the population growing as external pressure rises \cite{Sella:2000}. The latter aspect has been widely analyzed in the context of competing species \cite{gamere13,dodi13}, but has not received much attention in relation with the prevalence of altruism. 

Although general considerations about the prevalence of altruism in the context of the Public Goods Games can be inferred from the numerous studies on the topic \cite{ohhalino06,ko11}, the behavior of cooperation turns out to be very dependent on the specific dynamics considered \cite{ha06,rocusa09a}. This is the case when trying to evaluate the importance of the spatial heterogeneity and the formation of clusters of cooperators: Many studies \cite{namaiw97,iwnale98,lefedi03,thel03,lilihiiwboni13} explain the coexistence of cooperation and defection using the so called pair approximation, an approach that goes one step beyond mean field by tracking the dynamics of pairs of neighbors. However, pair approximation still assumes spatial homogeneity of the system. Hence, there is no need for the formation of clusters of cooperators for explaining their long-term survival. 

Recent works on the evolution of cooperation suggest the need of giving up on certain common statements of evolutionary game theory \cite{gaferutacusamo12,grrosemitr12,pejorawabosz17,sa18}. Particularly, some experiments on the dynamics of human cooperation show that people choose their strategy regardless the payoff of others \cite{grgrmisetrcumosa14}. Similar conclusions are given in the context of living beings \cite{doissi17}. See also recent experimental and numerical works on related topics \cite{waszpe12,kudi13,nadrfo16,leahha17}.

Here we study the evolution of cooperation in the framework of interacting particle systems. We model birth and death in a spatially extended population as a contact process and ask the following: What is the phase diagram of the contact process with an additional --- cooperative--- type of particle that supports survival at neighboring sites?
Our approach provides a natural framework to assess the effects of different mechanisms on the behavior of the system and on the survival of cooperativity, such as the dynamics of interactions, the fluctuation in the population size, the presence or absence of cooperation clusters, and the spatial variation of parameters, among others. 

The organization of the work is as follows. In Sec.~\ref{sec:2} we introduce the agent-based model of a population of cooperators and defectors living on a generic network. For later sections the main focus is on the square lattice, where the system has only three relevant parameters: the total number of sites $N$, the death parameter $p$, and the cost-of-altruism parameter $\epsilon$.  Section \ref{sec:3} includes stochastic simulations. We obtain the phase diagram in the parameter space $(p,\epsilon)$ showing the steady-state configurations of the system. The effect of $p$ being spatially dependent is also addressed. In Sec.~\ref{sec:4} the system is described theoretically. Three complementary formulations, using main-field or pair-approximation approaches, are given. They aim at describing the system under different physical conditions. Finally, a discussion and outlook of the main results are included in Sec.~\ref{sec:5}.

\section{Definition of the model}
\label{sec:2}

The model describes the evolution of a population on an arbitrary network with $N$ nodes. The set of neighbors of a node $i$ is denoted by $N_i$. The network is symmetric (undirected), so that $j \in N_i$ implies $i\in N_j$; also $i \notin N_i$ (no self-loops). Each agent in the population is either a cooperator $C$ or a defector $D$, with $c_i$ and $d_i$ being their respective numbers at node $i$. A site or node of the network holds at most one agent ($C$ or $D$) but it may also be empty ($E$), hence $0\le c_i+d_i\le 1$ and $c_id_i=0$. Thus the state of the system $S$ is given by 
\begin{equation}
\label{eq:1}
S=\{c_i,d_i\}_{i=1}^N\equiv \{x_i\}_{i=1}^N,\quad x_i\in\{c_i,d_i\}, 
\end{equation}
where $X$ is either a cooperator or a defector, and $x_i$ its number at site $i$. Moreover, the number $e_i=1-x_i$ gives $1$ if site $i$ is empty and $0$ if $x_i=1$. From condition $c_id_i=0$ we also have $e_ix_i=0$. 

A state transition is either the birth or the death of one agent at a site $i$. At the birth of a cooperator we set $c_i=1$ at a previously empty site $i$,
\begin{equation}
  e_i=1\xrightarrow{\pi_b(c_i,S)} c_i=1 
\end{equation}
Likewise for the birth of a defector, $d_i=1$ is set at an empty site $i$,
\begin{equation}
  e_i=1\xrightarrow{\pi_b(d_i,S)} d_i=1 
\end{equation}
These transitions occur at a rate proportional to the fraction of neighboring sites occupied by the agent type to be born, as \\
\begin{align}
\label{eq:2}
&\pi_b(c_i,S) = e_i \sum_{j \in N_i}c_j/k_j \equiv e_i \tilde{c}_i,& \\
\label{eq:3}
&\pi_b(d_i,S) = e_i \sum_{j \in N_i}d_j/k_j \equiv e_i \tilde{d}_i,&
\end{align}
where $k_i = |N_i|$ is the degree (number of those neighbors) of node $i$, and $\tilde{x}_i\equiv \sum_{j \in N_i} x_j/k_j$. The death of an agent is a state transition setting $c_i=0$ or $d_i=0$ at a previously occupied site $i$,
\begin{eqnarray}
  &&c_i=1\xrightarrow{\pi_d(c_i,S)} e_i=1, \\
  &&d_i=1\xrightarrow{\pi_d(d_i,S)} e_i=1,
\end{eqnarray}
with respective rates 
\begin{align}
  \nonumber 
  \pi_d(c_i,S)=&p\left[c_i\bar e_i+(1-\epsilon)c_i\bar c_i+(2-\epsilon)c_i\bar d_i\right]& \\
  \label{eq:4}
  =& pc_i\left\{1- \left[\bar c_i - (1-\epsilon) (\bar c_i+\bar d_i)\right] \right\},& \\
  \nonumber 
  \pi_d(d_i,S) =&p\left(d_i\bar e_i+d_i\bar d_i\right)& \\
  \label{eq:5}
  =&p d_i \left(1- \bar c_i\right),&
\end{align}
where now $\bar{x}_i\equiv k_i^{-1} \sum_{j \in N_i} x_j$. Agents die at a baseline rate $p$. This rate is reduced, however, by the fraction of adjacent sites occupied by a cooperator. The death rate of a cooperator, on the other hand, has an additional positive term proportional (with factor $1-\epsilon$) to the fraction of adjacent agents. This way, the parameter $\epsilon$ accounts for the cost of the altruistic act, the limit of $\epsilon=0$ corresponding to maximum cost where the altruist definitely loses its life for saving that of its neighbor. The other limit is costless altruism at $\epsilon=1$.

In the absence of cooperators, or in the absence of defectors with $\epsilon=0$, the model reduces to the contact process \cite{ha74,madi05} equivalent to the SIS (susceptible-infected-susceptible) model of epidemics \cite{he89,albrdrwu08}. The equivalence is obtained by mapping each empty site to a susceptible individual and each site with a defector to an infected individual.  

\section{Simulations}
\label{sec:3}

Let us first illustrate and numerically analyze the dynamics on periodic square lattices. As defined above, the model features non-ergodicity. Eventually both types of agents go extinct in a finite size system. In the simulations in this section, a slightly modified version of the model is employed: We set to zero the death rate of an agent currently being the only one of its type (C or D). This allows us to take long-term measurements of concentrations and distributions without having to restart the dynamics. Given the rates, simulations are performed with a standard Gillespie algorithm \cite{gi76,gi07}.

\subsection{Square lattice with homogeneous parameters} \label{subsec:sqlhom}

\begin{figure}
\centerline{\includegraphics[width=0.5\textwidth]{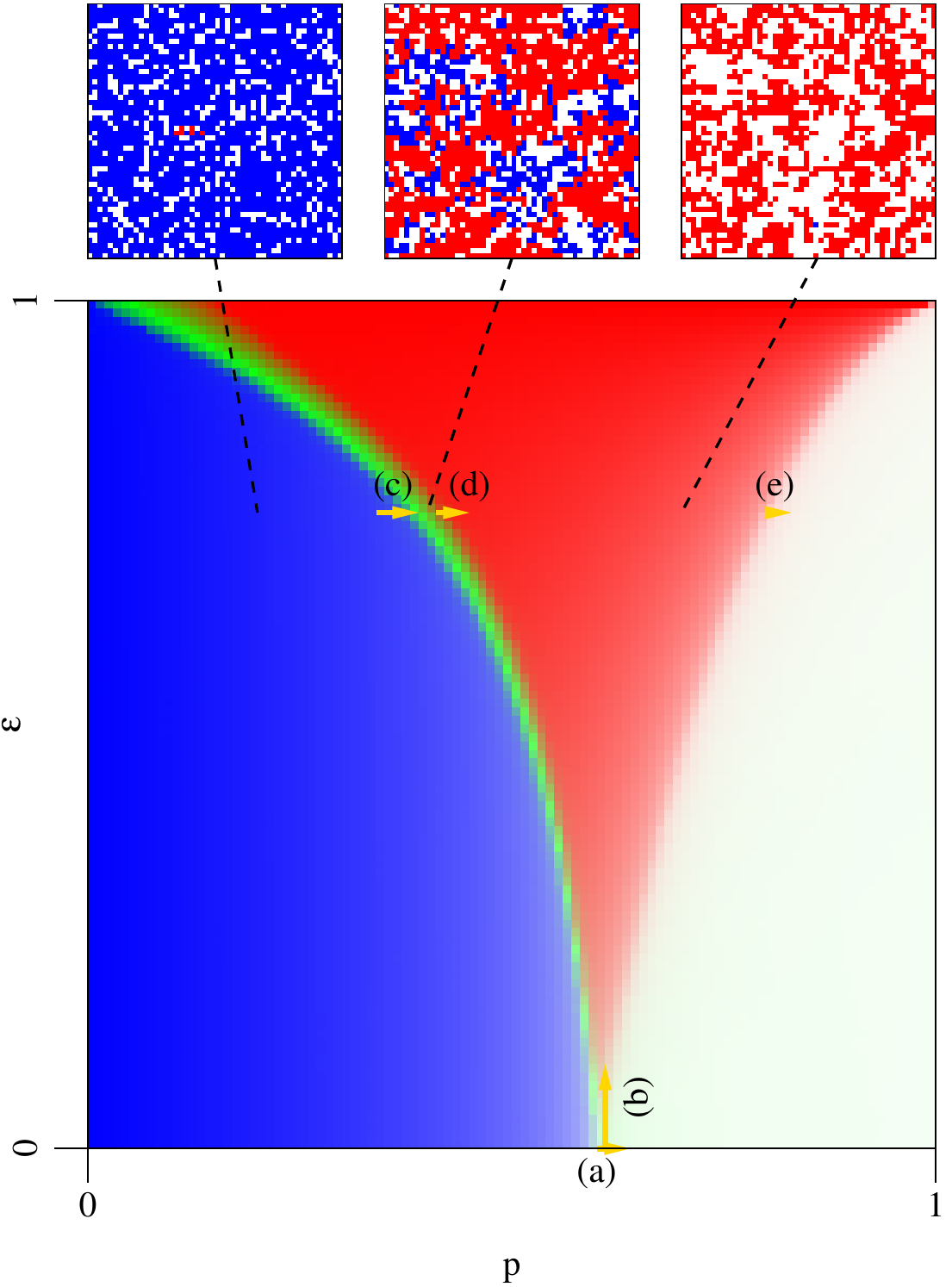}}
\caption{\label{fig:densities}
Average densities of agents on a square lattice of $50 \times 50$ sites as a function of parameters $p$ and $\epsilon$ (large center panel). For each combination of parameters, the agents' concentrations $\mean{c}$ and $\mean{d}$ are encoded by color. Red indicates high concentration of cooperators; blue indicates high concentration of defectors; green is for co-existence of the two types, and white for a low overall concentration of agents. Arrows labeled with letters (a)--(e) indicate parameter combinations further analyzed in Figure\ \ref{fig:distributions}. The panels in the top row are snapshots of typical system states encountered for $p \in \{0.2, 0.4, 0.7\}$ (panels left to right) with $\epsilon= 0.75$. 
}
\end{figure}

Figure \ref{fig:densities} shows the parameter dependence of the stationary mean concentrations of agents. At $\epsilon=0$, cooperators are absent in the whole range of $p$, while the concentration of defectors is positive for $p < p_c \approx 0.62$ and vanishes for $p>p_c$. Now fixing $0<\epsilon<1$ and increasing $p$ from $0$ to $1$, the concentration of defectors $\mean{d}$ still decreases with $p$. Before $\mean{d}$ reaches zero, however, the concentration of cooperators $\mean{c}$ becomes
positive. Simulations on square lattices of smaller size ($N=20^2$, $N=30^2$) and checks with $N=100^2$ yield results almost identical to those of Fig.\ \ref{fig:densities}.

\begin{figure*}
\centerline{\includegraphics[width=.8\textwidth]{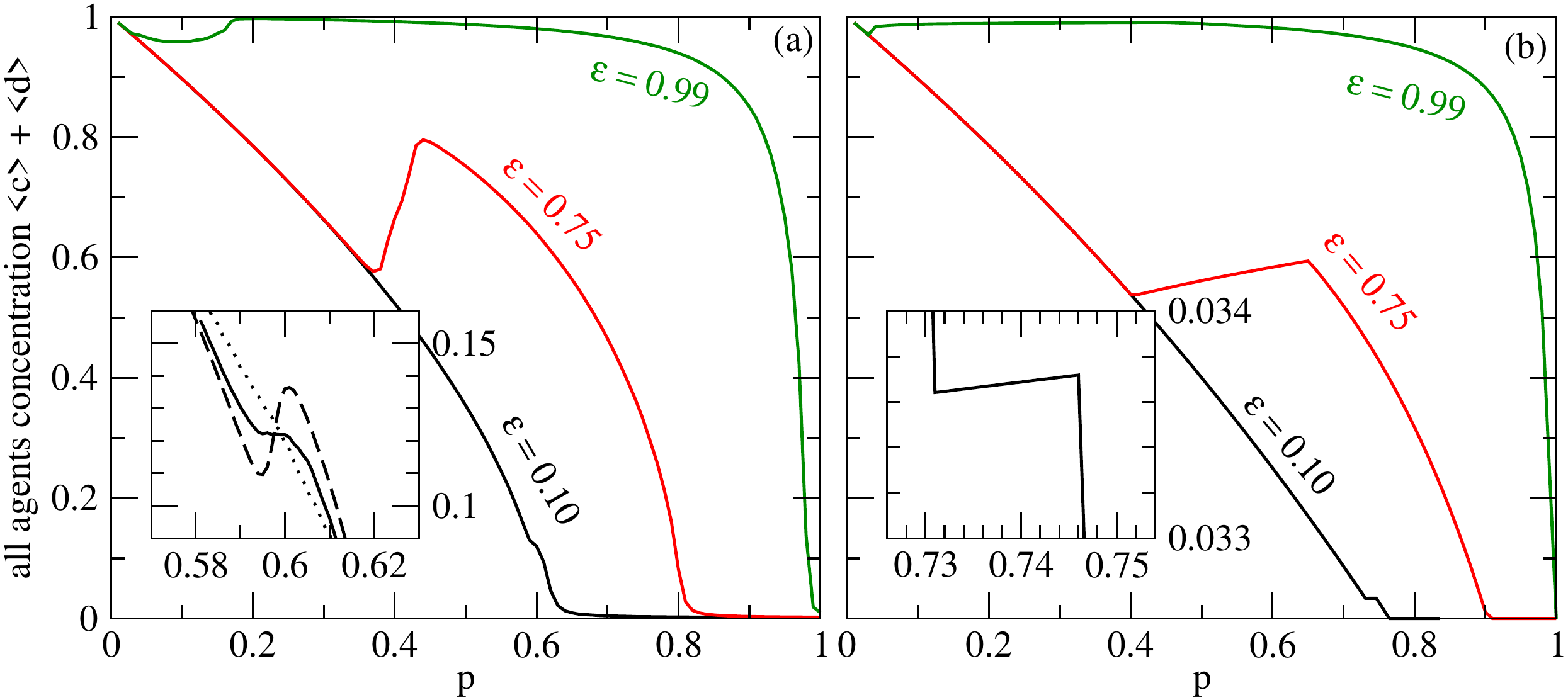}}
\caption{\label{fig:all_concentration}
Total concentration of agents (a) on square lattices with $N=50 \times 50$ sites and (b) from the numerical solution of the pair approximation, Eqs \eqref{eq:37}-\eqref{eq:41}. In both (a) and (b),
the three curves are for parameter values $\epsilon = 0.99, 0.75, 0.10$ (top to bottom). The insets zoom in on the curves for $\epsilon =0.10$. The inset of (a) shows these curves for different system sizes $N=30 \times 30$ (dotted curve), $N=50\times 50$ (solid curve), and $N=100 \times 100$ (dashed curve). 
}
\end{figure*}

In the coexistence regime of cooperators and defectors (green area in Figure \ref{fig:densities}), the growth of cooperation outweighs the decline of defection. Here the total concentration of agents grows with $p$,
\begin{equation}
\frac{\partial (\mean{c} + \mean{d})}{\partial p} >0~.
\end{equation}
Figure~\ref{fig:all_concentration}(a) explicitly shows this non-monotonicity by plotting $\mean{c} + \mean{d}$ versus $p$ for different choices of $\epsilon$.

\begin{figure*}
\centerline{\includegraphics[width=.8\textwidth]{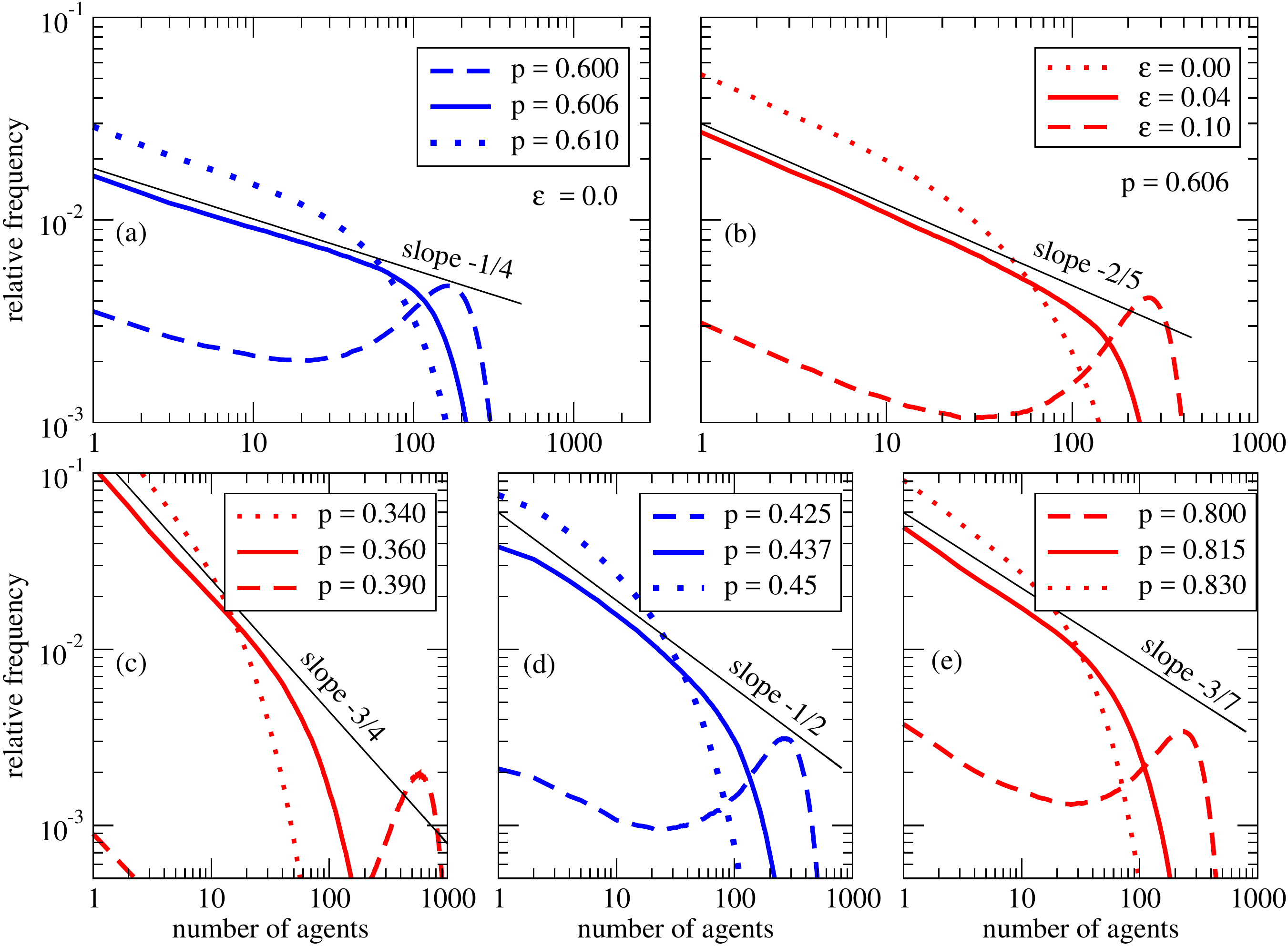}}
\caption{\label{fig:distributions}
Distributions of the number of agents on a square lattice with $50 \times 50$ sites. In the lower row, panels (c), (d), and (e), $\epsilon = 0.75$.
Each panel describes a transition between presence and absence of a type of agent. The transitions are also marked in Fig.\ \ref{fig:densities} with the panel identifiers (a)--(e).
} 
\end{figure*}

Let us now take a closer look at the transitions between the regimes observed in Figure~\ref{fig:densities}. To this end, we record the distributions in the number of agents (each type separately) and consider their changes under parameter variation. Figure~\ref{fig:distributions} shows this analysis for five transitions (a)-(e), also marked in the bottom panel of Figure~\ref{fig:densities}. 
Transitions in Figs.~\ref{fig:distributions}(a) and \ref{fig:distributions}(e) are extinctions of one type of agent in the absence of the other type. However, the transitions are distinguishable by the approximate exponents of the algebraic decay of distributions, giving $1/4$ for the extinction of defectors versus $3/7$ for cooperators. This indicates that, even in the absence of defectors and close to the extinction transition (e), the dynamics of cooperators is essentially different from the contact process.

Differences in the distributions of the order parameter (Figure~\ref{fig:densities}), however, do not contradict transitions (a)-(e) belonging to the same universality class. Transitions (a), (b) and (e) fulfill the premises of the directed percolation conjecture, cf.\ section 3.3.6 in \cite{hi00}. Transitions (c) and (d) do not fulfill the assumption of a unique absorbing state because only one type of
agent goes extinct at the transition. In preliminary numerical exploration (results not shown here), we have found the scaling of the order parameter (concentration of agents) compatible with the value $0.580(4)$ for exponent $\beta$ in directed percolation in two dimensions \cite{wazhligade13}. We conjecture that all transitions (a)-(e) belong to the universality class of directed percolation.

\subsection{Spatially dependent parameter $p$}

\begin{figure}
\centerline{\includegraphics[width=0.49\textwidth]{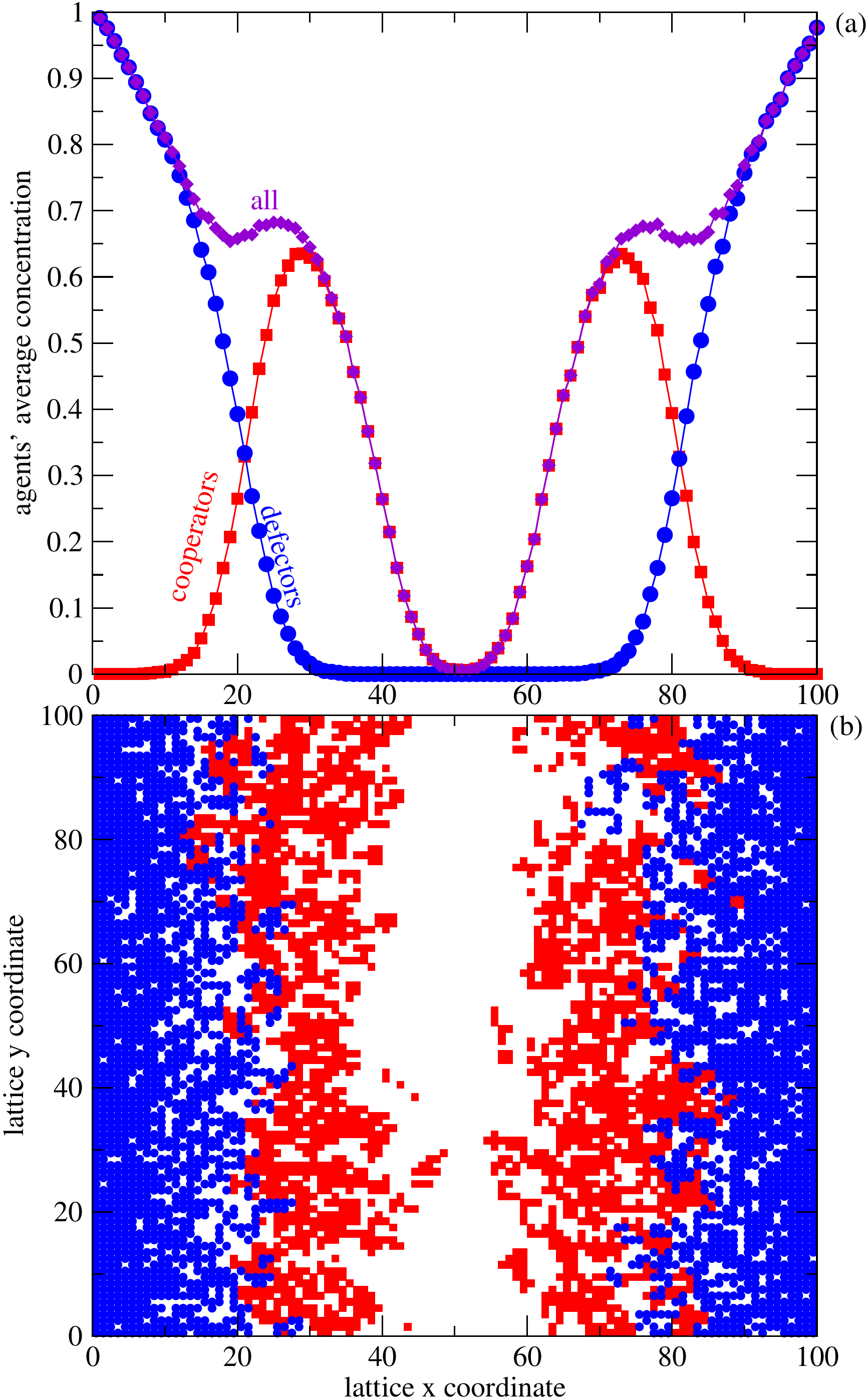}}
\caption{\label{fig:pspatial}
(a) Stationary mean concentrations for dynamics on a square lattice where parameter $p$ of the model varies with the horizontal location $x\in\{1,2,\dots,L\}$ according to Eq.\ (\ref{eq:p_of_x}). Lattice size is $L \times L$ with $L=100$. Cooperators' survival rate is $\epsilon = 0.75$ constant in space. Showing the concentration dependence on $x$, each plotted value is, for a given x, a uniform average over the $y$-coordinate of the lattice and over time $t\in [0,10 ^6]$.
(b) Snapshot of a state in the simulation as described for panel (a).
} 
\end{figure}

Let us study a variation of the model with a spatial dependence of the parameter $p$, a way of mimicking ecological conditions \cite{ch00,okle13}. For an agent at lattice site $(x,y)$, $x,y\in\{1,\dots,L\}$, the death rate is based on the parameter value
\begin{equation}\label{eq:p_of_x}
p(x) = \begin{cases}
\frac{2x-1}{L} & \text{if } x\le L/2 \\
\frac{2(L-x)+1}{L} & \text{otherwise.}
\end{cases}
\end{equation}
For $L$ even, the minimum value $1/L$ is assumed by $p(x)$ at $x=1$ and $x=L$; its maximum value $1-1/L$ is obtained at $x=L/2$ and $x=L/2+1$. The parameter $\epsilon$ remains spatially homogeneous, here $\epsilon =0.75$. 

Figure~\ref{fig:pspatial}(a) shows the concentration of agents as a function of lattice coordinate $x$, i.e. averaged over lattice coordinate $y$ and time. We see that the effect of parameter $p$ is local. The $p$-dependence of $\mean{c}$ and $\mean{d}$ observed under spatially homogeneous $p$ in Section~\ref{subsec:sqlhom} qualitatively matches that of the scenario with spatially dependent $p$.

\section{Analytic approximations}
\label{sec:4}

In this section, we derive three complementary theoretical descriptions of our model, defined in Sec.~\ref{sec:2}. The first two ones are based on a mean-field approximation, while the third one uses the pair approximation. As will be shown, the different approaches have different ranges of applicability and explain the prevalence/extinction and even the coexistence of altruism and defection under different physical and biological conditions. In the case of the pair approximation, a very similar phase diagram to the numerical one shown in Fig.~\ref{fig:densities} is obtained. 

Our starting point is the master equation for the probability $P(S,t)$ of finding the system in state $S$ at time $t$. By means of a probabilistic balance in the continuum time limit \cite{ka92}, and using the rates given by Eqs. \eqref{eq:2}-\eqref{eq:5}, the master equation reads as
\begin{equation}
  \label{eq:6}
  \begin{split}
    \partial_tP(S,t)=\sum_{i=1}^N&\sum_{x_i\in\{c_i,d_i\}}\left\{ (E^-_{x_i}-1)\left[\pi_b(x_i,S)P(S,t)\right] \right. \\
      & \left. +(E^+_{x_i}-1)\left[\pi_d(x_i,S)P(S,t)\right] \right\},
    \end{split}
\end{equation}
where the operators $E^\pm_{x_i}$ act on a generic function $f(x_1,\dots,x_i,\dots,x_N)$ as $E^\pm_{x_i}f(x_1,\dots,x_i,\dots,x_N)=f(x_1,\dots,x_i\pm 1,\dots,x_N)$, with $x_k\in\{c_k,d_k\}$, $k=1,\dots,N$.

By taking moments of the master equation \eqref{eq:6} we can derive equations for the mean numbers of cooperators and defectors in site $i$, $\mean{c_i}$ and $\mean{d_i}$. After using the relation $e_i=1-c_i-d_i$ and some manipulations, we obtain
\begin{eqnarray}
  \label{eq:7}
  \nonumber 
  \frac{d}{dt}\mean{c_i}=&& \mean{\pi_b(c_i)-\pi_d(c_i)}\\  
  \nonumber 
  =&& \mean{\tilde c_i e_i}-p\left[\mean{c_i\bar e_i}+(1-\epsilon)\mean{c_i\bar c_i}\right.\\  
  \nonumber 
  && \left.+(2-\epsilon)\mean{c_i\bar d_i}\right]\\  
  \nonumber 
  =&&-p\mean{c_i}+\mean{\tilde c_i}-\left[\mean{c_i\tilde c_i} -\epsilon p \mean{c_i\bar c_i} \right. \\
                         && \qquad \left. +p(1-\epsilon)\mean{c_i\bar d_i}+\mean{\tilde c_i d_i}\right], \\
  \label{eq:8}
  \nonumber 
  \frac{d}{dt}\mean{d_i}=&& \mean{\pi_b(d_i)-\pi_d(d_i)}\\  
  \nonumber =&& \mean{\tilde d_i e_i}-p\left[\mean{d_i\bar e_i}+\mean{d_i\bar d_i}\right]\\  
  \nonumber =&&-p\mean{d_i}+\mean{\tilde d_i} \\
                         &&-\left[\mean{c_i\tilde d_i} -p\mean{\bar c_i d_i}+\mean{d_i\bar d_i}\right], 
\end{eqnarray}
for $i=1,\dots,N$. Since the first moments are coupled to the second ones through correlations between neighbors, it is also convenient to derive equations for the two node correlations for neighboring sites, i.e. $\mean{x_ix_j}$ with $j\in N_i$:
\begin{eqnarray}
  \label{eq:9}
  \nonumber 
  \frac{d}{dt}\mean{c_ic_j}=&& \mean{c_i\pi_b(c_j)+\pi_b(c_i)c_j-c_i\pi_d(c_j)-\pi_d(c_i)c_j} \\
  \nonumber = && \mean{c_ie_j\tilde c_j}+\mean{\tilde c_i e_i c_j}-p\mean{c_ic_j(\bar e_i+\bar e_j)}\\
  \nonumber   && -p(1-\epsilon)\mean{c_ic_j(\bar c_i+\bar c_j)} \\
                 && -p(2-\epsilon)\mean{c_ic_j(\bar d_i+\bar d_j)}, \\
  \label{eq:10}
  \nonumber 
  \frac{d}{dt}\mean{c_id_j}=&& \mean{c_i\pi_b(d_j)+\pi_b(c_i)d_j-c_i\pi_d(d_j)-\pi_d(c_i)d_j} \\
  \nonumber =&& \mean{c_ie_j\tilde d_j}+\mean{\tilde c_i e_i d_j}-p\mean{\bar e_ic_id_j}\\
  \nonumber && -p\mean{c_id_j(\bar e_j+\bar d_j)}-p(1-\epsilon)\mean{\bar c_ic_id_j} \\
            && -p(2-\epsilon)\mean{\bar d_ic_id_j}, \\
  \label{eq:11}
  \nonumber 
  \frac{d}{dt}\mean{d_id_j}=&& \mean{d_i\pi_b(d_j)+\pi_b(d_i)d_j-d_i\pi_d(d_j)-\pi_d(d_i)d_j} \\
  \nonumber = && \mean{d_ie_j\tilde d_j}+\mean{\tilde d_i e_i d_j}-p\mean{d_id_j(\bar e_i+\bar e_j)}\\
                            && -p\mean{d_id_j(\bar d_i+\bar d_j)},
\end{eqnarray}
where $\tilde x_i$ and $\bar x_i$ are defined just after Eqs. \eqref{eq:3} and \eqref{eq:5}, respectively. The two remaining moments, $\mean{c_ie_j}$ and $\mean{d_ie_j}$ can be obtained from the previous ones by means of the identity $1=e_i+c_i+d_i$, as $\mean{c_ie_j}=\mean{c_i}-\mean{c_ic_j}-\mean{c_id_j}$ and $\mean{d_ie_j}=\mean{d_i}-\mean{d_id_j}-\mean{d_ic_j}$.

Although the system of Eqs. \eqref{eq:7}-\eqref{eq:11} are exact and valid for any structure of neighbors (network), it is not closed, due to the presence of three nodes correlations. Therefore, in order to have a closed set of equations, three approximations are explored. The first two ones make use of the mean-field approximation, where two node correlations are ignored, and the third one uses pair approximation. Furthermore, we restrict ourselves to regular networks where $k_i=k$ for all $i$, so as to simplify the description (now $\tilde x_i=\bar x_i=k^{-1} \sum_{j \in N_i} x_j$). 

\subsection{Exact relations}
Before proceeding with the approximations, some exact relations will be derived. They apply for homogeneous steady-state configurations. 

Consider first the case of only defectors. Since $\mean{c}=0$, we also have $\mean{cc}=\mean{cd}=\mean{ce}=0$. Using Eq.~\eqref{eq:8}, together with $\mean{de}=\mean{d}-\mean{dd}$, we have 
\begin{equation}
  \label{eq:42}
  \mean{dd}=(1-p)\mean{d},
\end{equation}
and, with Eq.~\eqref{eq:11} and the identity $\mean{dde}+\mean{ddd}=\mean{dd}$,
\begin{equation}
  \label{eq:43}
  p\left[1-k(1-p)\right]\mean{d}+(k-1)\mean{ded}=0.
\end{equation}
which implies, in order to have positive solutions, $1-k(1-p)\le 0$, that is
\begin{equation}
  \label{eq:44}
  p\le 1-\frac{1}{k}. 
\end{equation}
This is an overestimation of the extinction probability of defectors, for all $\epsilon\in[0,1]$. For $\epsilon=0$, where the model is the SIS model, and the square lattice ($k=4$), the previous estimation is $0.75$ while the one from the simulations is around $0.62$ \cite{saol02, vofama09}, see also Fig.\ \ref{fig:densities}.

For the only-cooperator case, it is $\mean{d}=0$ and $\mean{dd}=\mean{cd}=\mean{de}=0$. Using Eq.~\eqref{eq:7} together with $\mean{ce}=\mean{c}-\mean{cc}$, we get 
\begin{equation}
  \label{eq:55}
  \mean{cc}=\frac{1-p}{1-\epsilon p}\mean{c},
\end{equation}
and, with Eq.~\eqref{eq:9} and the identity $\mean{cce}+\mean{ccc}=\mean{cc}$,
\begin{equation}
  \label{eq:56}
  \begin{split}
    &\frac{p}{1-\epsilon p}\left[p(1-\epsilon)-(k-1)(1-p)\right]\mean{c}\\
    &\qquad +(k-1)\left(\mean{cec}+p\epsilon\mean{ccc}\right)=0,
  \end{split}
\end{equation}
which now implies $p(1-\epsilon)-(k-1)(1-p)\le 0$ or
\begin{equation}
  \label{eq:57}
  p\le 1-\frac{1-\epsilon}{k-\epsilon}\ge 1-\frac{1}{k}. 
\end{equation}
Again, this is an overestimation of the critical probability extinction when there are only cooperators in the system. The critical value here is bigger or equal to the one of Eq. \eqref{eq:44}, as expected due to the altruistic benefit. Equation \eqref{eq:57} also provides an estimation of the dependence of the critical probability on $\epsilon$. In particular, it tends to $1$ for $\epsilon\to 1$, in agreement with the numerical simulations of Fig. \ref{fig:densities}.

Equations \eqref{eq:44} and \eqref{eq:57}, and also the other relations, are the same for $\epsilon=0$ provided we interchange the types of particles, because the model with only defectors and only cooperators coincide in this limit. This can be seen from the rates defining the dynamics in Eqs. \eqref{eq:2}-\eqref{eq:5}: The rates for defectors in the absence of cooperators are the same as the rates for cooperators in the absence of defectors at $\epsilon=0$. 

\subsection{Global mean-field approximation} \label{sec:global_mean}

For the global mean-field case, equivalent to the dynamics on a complete graph in the limit of infinite system size, correlations among nodes are absent. In general, assuming the mean-field approximation implies the following two approximations:
\begin{eqnarray}
  \label{eq:12}
  && \mean{x_i x_j}\simeq \mean{x_i}\mean{x_j}, \qquad i\ne j \\
  \label{eq:13}
  && \mean{x_i}\simeq \mean{x_j} \equiv \mean{x}, \qquad \text{for all } i.
\end{eqnarray}
This is also a good approximation when there is no correlation expected between the agents, for instance when there is one kind of agent and the distribution of empty sites is homogeneously distributed. Then, the concentrations $\mean{c}$ and $\mean{d}$ of cooperators and defectors evolve, according to Eqs.~\eqref{eq:7} and \eqref{eq:8}, as 
\begin{eqnarray}
  \label{eq:14}
  \nonumber
  \frac{d}{dt}\mean{c}=&&\mean{c} \left\{(1-p)-(1-\epsilon p)\mean{c} \right. \\
                        && \qquad \left.-\left[1+p(1-\epsilon)\right]\mean{d}\right\}, \\  
  \label{eq:15}
   \frac{d}{dt}\mean{d}=&&\mean{d} \left[(1-p)-(1-p)\mean{c}-\mean{d}\right].
\end{eqnarray}

The system \eqref{eq:14} and \eqref{eq:15} can be used now to analyze the homogeneous steady-state solutions. Requiring stationarity, $\frac{d}{dt}\mean{c}=\frac{d}{dt}\mean{d}=0$, we find the trivial solution $\mean{c}=\mean{d}=0$ (all sites empty) and, two other, nontrivial ones, namely
\begin{eqnarray}
  \label{eq:16}
  && \mean{c}=0\; \mathrm{and} \; \mean{d}=1-p, \\
  \label{eq:17}
  && \mean{c}=\frac{1-p}{1-\epsilon p}\; \mathrm{and} \; \mean{d}=0.
\end{eqnarray}

The trivial solution is clearly unstable, since the coefficient $1-p$ of the less degree terms in Eqs. \eqref{eq:14} and \eqref{eq:15} is positive for $p< 1$. However, it is an absorbing state, and their presence becomes important for small system sizes, as already mentioned in Sec. \ref{sec:3}. 

In order to assess the stability of the solution with only defectors, consider the perturbation of Eq.~\eqref{eq:16}: $\mean{c}=0+\mean{c}_1$ and $\mean{d}=1-p+\mean{d}_1$ with $\mean{c}_1\sim \mean{d}_1$. Then, up to linear order in the perturbations, we have
\begin{eqnarray}
  \label{eq:18}
  &&\frac{d}{dt}\mean{c}_1\simeq -p(1-p)(1-\epsilon)\mean{c}_1, \\
  \label{eq:19}
  &&\frac{d}{dt}\mean{d}_1\simeq -(1-p)\left[(1-p)\mean{c}_1+\mean{d}_1\right].
\end{eqnarray}
The first equation, and hence the second one, have $\mean{c}_1=\mean{d}_1=0$ as the steady solution, revealing the stable character of \eqref{eq:16}. Proceeding similarly with the only-cooperators solution, Eq. \eqref{eq:17}, we obtain the system 
\begin{eqnarray}
  \label{eq:20}
  \frac{d}{dt}\mean{c}_1\simeq && -\frac{1-p}{1-\epsilon p}\left\{(1-\epsilon p)\mean{c}_1 \right.\\ 
    && \nonumber \qquad \left. +\left[1+p(1-\epsilon)\right]\mean{d}_1\right\}, \\  
  \label{eq:21}
  \frac{d}{dt}\mean{d}_1\simeq && p(1-p)\frac{1-\epsilon}{1-\epsilon p}\mean{d}_1,
\end{eqnarray}
which now reveals the unstable character of the solution, since the solution of Eq.~\eqref{eq:21} increases exponentially with time. According to this analysis, in well-mixed populations, cooperators go extinct. 

\subsection{Local mean-field approximation}
We can go one step beyond the global mean-field approximation by considering situations where the concentrations of cooperators and defectors change from site to site. In particular, we suppose situations where the site dependence can be encoded through a vector $\mathbf r$, which is nothing but the vector of space position in a regular graph. This way, we deduce in the sequel a macroscopic description that removes one of the approximation of the global mean field, namely that of Eq.~\eqref{eq:13}, but still neglects correlations, Eq.~\eqref{eq:12}. The procedure is similar to the one used in Ref. \cite{khlohe17}. 

By looking at the dynamics on a length scale $L$ much larger than the typical distance between sites $l$, the relevant quantities become local concentrations:
\begin{eqnarray}
  \label{eq:22}
  && \kappa(\mathbf r) \equiv \mean{c_i}, \\
  \label{eq:23}
  && \delta(\mathbf r) \equiv \mean{d_i}.
\end{eqnarray}
In a regular graph in $\mathbb R^d$, for example, $\kappa(\mathbf r)$ and $\delta(\mathbf r)$ give the number of cooperators and defectors inside a region of volume $l^d$ centered at position $\mathbf r$. The new quantities are assumed to be smooth functions of $\mathbf r$, a property that allows us to relate any density of site $j\in N_i$ and position $\mathbf l$, say $\chi(\mathbf r+\mathbf l)=\kappa(\mathbf r+\mathbf l)$ or $\chi(\mathbf r+\mathbf l)=\delta(\mathbf r+\mathbf l)$ at position $\mathbf r$, with that of site $i$, $\chi(\mathbf r)$, as
\begin{eqnarray}
  \label{eq:24}
  &&\chi(\mathbf r+\mathbf l)\simeq \chi(\mathbf r)+\nabla \chi(\mathbf r)\cdot \mathbf l+\frac{1}{2}\nabla \nabla \chi(\mathbf r): \mathbf l \mathbf l,
\end{eqnarray}
Hence, we have 
\begin{equation}
  \label{eq:25}
  \mean{\bar x_i}=\frac{1}{k_i}\sum_{k\in N_i}\chi(\mathbf r+\mathbf l_k) \simeq  \chi(\mathbf r)+\nabla^2_r\chi(\mathbf r).
\end{equation}
where we have assumed $\sum_{k\in N_i}\mathbf l_k\simeq 0$, which is an exact expression for a regular square lattice and quiet a good approximation for \emph{isotropic} configurations. Moreover, 
\begin{equation}
  \label{eq:26}
\nabla^2_r\chi(\mathbf r)\equiv\frac{1}{2k_i}\sum_{k\in N_i}\nabla \nabla \chi(\mathbf r): \mathbf l_k \mathbf l_k\simeq \frac{l^2}{2d}\nabla^2\chi(\mathbf r),
\end{equation}
which is valid, again, under \emph{isotropic} configurations of sites.

With approximations \eqref{eq:12}, \eqref{eq:25}, and \eqref{eq:26}, the exact system \eqref{eq:7} and \eqref{eq:8} becomes the following reaction-diffusion system
\begin{eqnarray}
  \label{eq:27}
  \nonumber &&\partial_t\kappa=\kappa\left\{(1-p)-(1-\epsilon p)\kappa-\left[1+p(1-\epsilon)\right]\delta \right\} \\
            && \quad +\left[1-(1-\epsilon p)\kappa-\delta\right]\nabla_r^2\kappa -p(1-\epsilon)\kappa \nabla_r^2\delta, \\
  \label{eq:28}
  \nonumber &&\partial_t\delta=\delta\left\{(1-p)-(1-p)\kappa-\delta \right\}  \\
            && \qquad +\left[1-\kappa-\delta\right]\nabla_r^2\delta+p\delta \nabla_r^2\kappa.
\end{eqnarray}
As expected, we recover the mean-field description for homogeneous solutions, hence we still have the solutions given in Eqs. \eqref{eq:16} and \eqref{eq:17}. However, an important benefit of the present description, if compared to that of the global mean-field approximation, is the possibility of studying the latter solutions under local perturbations, in contrast to homogeneous and global ones done in the previous subsection. 

Consider the homogeneous solution of Eq.~\eqref{eq:16}, $\kappa_0=0$ and $\delta_0=1-p$. Following the standard linear stability analysis, we seek solutions of the form $\kappa=\kappa_0+\kappa_1$ and $\delta=\delta_0+\delta_1$, with $\kappa_1\sim \delta_1\ll \delta_0$. After linearizing and seeking solutions of the form $\chi_1=\tilde \chi_1 e^{i\boldsymbol \xi \cdot \mathbf r}$, system \eqref{eq:27} and \eqref{eq:28} becomes
\begin{eqnarray}
  \label{eq:29}
  &&\partial_t\tilde\kappa_1=-p\left[(1-p)(1-\epsilon)+\frac{l^2}{2d}\xi^2\right]\tilde\kappa_1, \\
  \label{eq:30}
  &&\partial_t\tilde\delta_1=-\left[(1-p)+p\frac{l^2}{2d}\xi^2\right]\left[(1-p)\tilde\kappa_1+\tilde\delta_1\right].
\end{eqnarray}
The steady state solution for any wavelength $\mathbf \xi$ is the trivial one, meaning that the solution of only defectors is linearly stable: any initial and small spatial perturbation in the number of defectors (and also cooperators) tends to zero as time increases.

Proceeding similarly with the solution of Eq.~\eqref{eq:17}, we get 
\begin{eqnarray}
  \label{eq:31}
  \nonumber &&\partial_t\tilde\kappa_1=-\left[(1-p)(1-\epsilon)+p\frac{l^2}{2d}\xi^2\right]\tilde\kappa_1, \\
            && \qquad -\frac{1-p}{1-\epsilon p}\left[1+p(1-\epsilon)\left(1-\frac{l^2}{2d}\xi^2\right)\right]\tilde\delta_1, \\
  \label{eq:32}
            &&\partial_t\tilde\delta_1=\frac{p(1-\epsilon)}{1-\epsilon p}\left[(1-p)-\frac{l^2}{2d}\xi^2\right]\tilde\delta_1.
\end{eqnarray}
In this case, the stability of the system depends on the value of $\xi$. Setting $\xi=2\pi/L$, the smallest allowed value for the given boundary conditions, the solution \eqref{eq:17} is stable for $p<p_c^*$ with
\begin{equation}
  \label{eq:33}
  p_c^*=1-\frac{2\pi^2l^2}{dL^2}\simeq 1-\frac{2\pi^2}{dN^{\frac{2}{d}}},
\end{equation}
where we have used the approximation $L/l\simeq N^{1/d}$. This means that, under this approximation, the only-cooperators solution is stable for systems small enough. For $N\to \infty$ it is $p_c^*\to 1$, and the solution is always unstable, and we recover the result of mean field. 

Although the local mean-field approximation could in principle be seen as very crude, it shows the importance of taking into account the system size while describing altruism, as already pointed out in Ref. \cite{sa18}. In this case, the inclusion of spatial dependence, while still neglecting correlations, stabilizes the only-cooperators solution for $p<\tilde p_c$. Moreover, the results suggest the existence of other solutions, spatially non-homogeneous ones, and the possibility of discontinuous (first-order) transitions among them. This is because the only-defectors solution keeps always linearly stable, with no other stable solution close to it. 

\begin{figure}[ht]
\centerline{\includegraphics[width=0.49\textwidth]{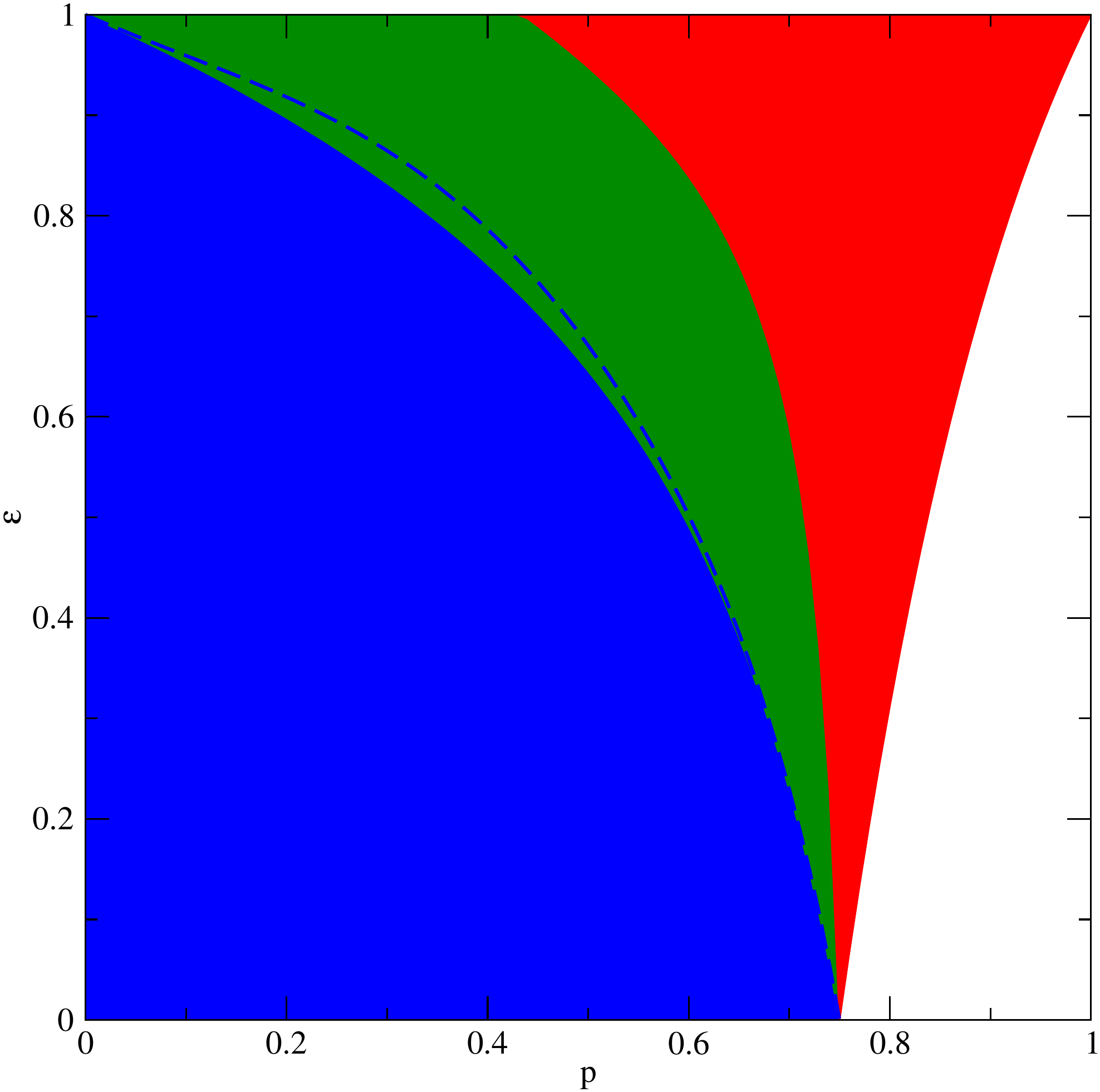}}
\caption{ \label{fig:pair_app_analytic}
Phases in pair approximation. The extinction transition between the regime of only cooperators (red area) to the empty system (white area) is given by the expression in $p$ and $\epsilon$ in Equation~\eqref{eq:67}. The transition between coexistence (green area) and the regime of only cooperators is described by Equation~\eqref{eq:78} using expressions \eqref{eq:79}-\eqref{eq:81}.
The transition between coexistence and the regime of only defectors (blue area) has an approximate description (dashed curve) in Equation~\eqref{eq:82} using expressions \eqref{eq:83}-\eqref{eq:85}. The exact solution (boundary between blue and green area) has been obtained as well, details given elsewhere.}
\end{figure}

\subsection{Pair approximation}

The previous mean-field approaches are expected to fail when the concentration of defectors and cooperators are locally correlated. Since births occur among neighboring sites, correlations are expected to be important, in general. Hence, we reconsider system \eqref{eq:7}-\eqref{eq:11}, and try to express the three nodes moments as a function of the one and two nodes mean values. Although different approaches are possible (see for instance \cite{khlohe17}), we explore here the so-called pair approximation. Pair approximation has been extensively applied to a variety of stochastic processes defined on a network, aiming at describing different situations as diverse as spin dynamics \cite{olmesa93,gl13}, opinion dynamics \cite{vaeg08,scbe09,vacasa10,pecasato18,pecasato18a}, epidemics \cite{ledu96,eake02,tasigrhoki12}, and population dynamics \cite{namaiw97,iwnale98,lilihiiwboni13,lefedi03,thel03}. In each of the cases, the pair approximation assumes that the probability of a given node quantity $x_i$ conditioned to the values of a neighboring site $x_j$ and to a next-neighboring site $x_k$ is independent of the latter \cite{maogsasa92}: $\text{Prob}(x_i|x_jx_k)\simeq \text{Prob}(x_i|x_j)$. In other words, the state of a neighbor of a given node is considered to be independent of the state of another neighbor. In our model, where $x_i\in\{c,d,e\}$ takes the values $0$ or $1$, the mean values $\mean{x_i x_j}$ and $\mean{x_i x_j x_k}$ are essentially the respective probabilities of the given quantities, hence, under the pair approximation $\mean{x_ix_jx_k}=\text{Prob}(x_i|x_jx_k)\text{Prob}(x_jx_k)\simeq P(x_i|x_j)\text{Prob}(x_jx_k)$, we have
\begin{equation}
  \label{eq:36}
  \mean{x_ix_jx_k}\simeq\frac{\mean{x_ix_j}\mean{x_jx_k}}{\mean{x_j}}.
\end{equation}
Note that the order of appearance of the variables inside the brackets is important: $x_i$ refers to a node which is a neighbor of $x_j$ and $x_j$ is a neighbor of $x_k$. Observe that the pair approximation keeps the correlations regardless of the occupancy of the middle node, namely $\sum_{x_j\in\{c,d,e\}}\mean{x_ix_jx_k}=\mean{x_i1x_k}\simeq\sum_{x_j\in\{c,d,e\}}\frac{\mean{x_ix_j}\mean{x_jx_k}}{\mean{x_j}}\ne \mean{x_i}\mean{x_j}$, in general.

For simplicity, we consider homogeneous situations for which system \eqref{eq:7}-\eqref{eq:11}, within the pair approximation of Eq. \eqref{eq:36}, becomes
\begin{eqnarray}
  \label{eq:37}
  \nonumber
  \frac{d}{dt}\mean{c}&=&(1-p)\mean{ce}-p(1-\epsilon)\mean{cc}\\ 
                       && -p(2-\epsilon)\mean{cd}, \\
  \label{eq:38}
  \frac{d}{dt}\mean{d}&=&(1-p)\mean{de}-p\mean{dd}, \\
  \label{eq:39}
  \nonumber
  \frac{k}{2}\frac{d}{dt}\mean{cc}&=&\mean{ce}-p(1-\epsilon)\mean{cc}\\
  \nonumber  &&+(k-1)\left\{\frac{\mean{ce}^2}{\mean{e}}-p\left[\frac{\mean{cc}\mean{ce}}{\mean{c}} \right. \right. \\
                       &&\left.\left.+(1-\epsilon)\frac{\mean{cc}^2}{\mean{c}}+(2-\epsilon)\frac{\mean{cc}\mean{cd}}{\mean{c}}\right]\right\}, \\
  \label{eq:40}
  \nonumber  k\frac{d}{dt}\mean{cd}&=&-p(2-\epsilon)\mean{cd}+(k-1)\left\{2\frac{\mean{ce}\mean{ed}}{\mean{e}} \right. \\
  \nonumber  &&-p\left[\frac{\mean{ec}{\mean{cd}}}{\mean{c}}+\frac{\mean{cd}\mean{de}}{\mean{d}}+\frac{\mean{cd}\mean{dd}}{\mean{d}} \right. \\
                      &&\left.\left.+(1-\epsilon)\frac{\mean{cc}\mean{cd}}{\mean{c}}+(2-\epsilon)\frac{\mean{cd}^2}{\mean{c}}\right]\right\}, \\
  \label{eq:41}
  \nonumber  
  \frac{k}{2}\frac{d}{dt}\mean{dd}&=&\mean{de}-p\mean{dd}+(k-1)\left[\frac{\mean{de}^2}{\mean{e}} \right.\\
                      &&\left. -p\left(\frac{\mean{dd}\mean{de}}{\mean{d}}+\frac{\mean{dd}^2}{\mean{d}}\right)\right],
\end{eqnarray}
where $\mean{xy}$ is for any two adjacent nodes with particles $x$ and $y$. Hence, $\mean{xy}=\mean{yx}$. 

\subsubsection{Steady-state solutions}

The system \eqref{eq:37}-\eqref{eq:41}  has several steady-state solutions. The most obvious one it the trivial solution, without particles, $\mean{c}=\mean{d}=0$. This is the absorbing state we have already mentioned.

By setting all time derivatives of Eqs. \eqref{eq:37}-\eqref{eq:41}  to zero and $c=0$, we obtain the steady-state solution for defection only:
\begin{equation}
  \label{eq:53}
  \mean{d}=\frac{(1-p)k-1}{k-(1+p)},
\end{equation}
valid for
\begin{equation}
  \label{eq:54}
  p\le 1-\frac{1}{k}.
\end{equation}
Observe that the previous inequality is the same as the one in Eq. \eqref{eq:44}, derived using exact relations. However, in this case, $p=1-1/k$ is the exact critical value for the extinction of defectors in the absence of cooperators, within the pair approximation. 

The only-cooperators solution is obtained from Eqs. \eqref{eq:37}-\eqref{eq:41}  as a steady-state solution with $d=0$. Now,
\begin{equation}
  \label{eq:66}
  \mean{c}=\frac{(1-p)k-(1-\epsilon p^2)}{[k-(1+p)](1-\epsilon p)},
\end{equation}
for 
\begin{equation}
  \label{eq:67}
  p\le  \frac{k-\sqrt{k^2-4\epsilon(k-1)}}{2\epsilon}\ge 1-\frac{1}{k}.
\end{equation}
The equality of the last relation holds for $\epsilon \to 0$. For the other limiting value of $\epsilon$, i.e. $\epsilon \to 1$, the upper allowed value of $p$ is $1$, as we also obtained exactly.

Other steady-state solutions describing coexistence, but close to the previous ones, can also be found as follows. First, we notice that the system \eqref{eq:37}-\eqref{eq:41}, under the steady-state condition, can be reduced to a nonlinear system of only two equations with $\mean{c}$ and $\mean{d}$ as unknown quantities. Second, we seek solutions close to the one-type ones, i.e. $\mean{c}\simeq \frac{1-k(1-p)-\epsilon p^2}{(1-k+p)(1-\epsilon p)}$ and $\mean{d}\simeq 0$ for the only-cooperators case and $\mean{c}\simeq 0$ and $\mean{d}\simeq \frac{1-k(1-p)}{1-k+p}$ for the only-defectors. For the former case, the resulting equations are linear and nontrivial solutions appear below the following line:
\begin{equation}
  \label{eq:78}
  \epsilon_c(p)=\frac{A(p)B(p)}{C(p)+\sqrt{C^2(p)-A^2(p)B(p)}},
\end{equation}
with 
\begin{eqnarray}
  \label{eq:79}
  A(p)&=&2p(1-p)(k-1-kp) \\
  \label{eq:80}
  B(p)&=&(k+1-(k+2)p+2p^2)/[p(1-p)], \\
  \nonumber
  C(p)&=&k(k-1)-(2k^2-3k+2)p \\
  \label{eq:81}
      &&+(k^2-3k+1)p^2+(k+2)p^3-2p^4.
\end{eqnarray}
For the only-defectors case, the resulting set of equation is nonlinear, but one can still find a condition for a nontrivial solution to exist. Now, the nontrivial solutions are above $\epsilon_d(p)$, which has the following approximate expression
\begin{equation}
  \label{eq:82}
  \epsilon_d(p)\simeq \frac{E(p)}{F(p)}\left[1-\sqrt{1-\frac{2G(p)F(p)}{E^2(p)}}\right],
\end{equation}
with
\begin{eqnarray}
  \nonumber 
  E(p)&=&(k-1)^3+(k-1)(4k^2-7k+8)p\\ 
  \nonumber 
      && -\{k[5k(k-5)+28]-3\}p^2\\ 
  \label{eq:83}
      && -(13k-5)kp^3, \\
  \nonumber 
  F(p)&=&2p\{(k-1)(2k^2-3k+2)\\
  \nonumber 
      &&\quad -(2k^3-14k^2+14k-1)p \\
  \label{eq:84}
      && \qquad -(9k-4)kp^2\}, \\
  \nonumber 
  G(p)&=&(k-1-kp)[(k-1)^2\\
  \label{eq:85}
      &&\quad +(3k^2-7k+10)p+2(3k-1)p^2].
\end{eqnarray}

Since $\epsilon_c(p)\ge \epsilon_d(p)$, the coexistence solutions are in the region in between the two lines, as shown in Fig.\ \ref{fig:pair_app_analytic} for the square lattice ($k=4$).

Finally, there may be other solutions describing coexistence not necessarily close to the only-cooperator nor only-defectors ones. This can be shown explicitly for $\epsilon=1$, for which we can obtain explicit expressions. After some algebra, we get
\begin{eqnarray}
  \label{eq:68}
  && \mean{c}=\frac{2(k-1)(2k-3)(1-3p)}{(1-p)[4(k-1)(k-p-2)+p+1]}, \\
  \label{eq:69}
  && \mean{d}=\frac{(2k-3)[(4k-5)p-1]}{4(k-1)(k-p-2)+p+1}, \\
  \label{eq:70}
  && \mean{cd}=\frac{4(k-1)(k-p-2)+p+1}{2(k-1)(2k-3)}\mean{c}\mean{d},
\end{eqnarray}
valid for 
\begin{equation}
  \label{eq:71}
  \frac{1}{4k-5}\le p\le \frac{1}{3}.
\end{equation}
Moreover, it can be seen that this solution is linearly unstable, with only one unstable mode. However, the characteristic time of the unstable mode is much slower than the others, meaning that the system can stay close to the solution for a long time. 

\subsubsection{Stability of the steady-state solutions}
The stability of the only-defectors and only-cooperators solutions have been studied by means of a modified linear stability analysis of system \eqref{eq:37}-\eqref{eq:41}, following several steps. First, using the identities $\mean{ce}=\mean{c}-\mean{cc}-\mean{cd}$ and $\mean{de}=\mean{d}-\mean{cd}-\mean{dd}$, all mean values are expressed in terms of $\mean{c}$, $\mean{d}$, $\mean{cc}$, $\mean{cd}$, and $\mean{dd}$. Second, the homogeneous solution is linearly perturbed as 
\begin{equation}
  \label{eq:86}
  \mathbf  u=\mathbf  u_0+\gamma \mathbf  u_1,
\end{equation}
with $\mathbf  u=(\mean{c},\mean{d},\mean{cc},\mean{cd},\mean{dd})$ the vector of the homogeneous solutions, $\mathbf  u_0$ is the vector of the unperturbed solutions, $\mathbf  u_1$ is the perturbation vector, and $\gamma$ a perturbative parameter. Third, the proposed solution is replaced in \eqref{eq:37}-\eqref{eq:41} and the resulting system is expanded up to linear order in $\gamma$. Contrary to the usual linear perturbation schemes, we obtain a nonlinear closed system of equations for the unknown perturbation quantities $u_{1,i},$ for $i=1,\dots, 5$. For both, the only-cooperators and only-defectors solutions, the equation for the perturbation can be written as
\begin{equation}
  \label{eq:87}
  \frac{d}{dt}\mathbf  u_1=M(\beta)\mathbf  u_1,
\end{equation}
with $M$ being a matrix and $\beta$ a linear function of $\mean{cc}/\mean{c}$, $\mean{cd}/\mean{c}$,  $\mean{cd}/\mean{d}$, and $\mean{dd}/\mean{d}$, whose explicit form depends on the solution considered. In any case, $\beta$ is a bounded function, since $0\le \mean{xy}/\mean{x}\le 1$ for $x,y\in\{c,d\}$. Finally, the asymptotic behavior of $\mathbf u_1(t)$ for $t\to \infty$, hence the stable or unstable character of $\mathbf  u_0$, can be determined from the spectra of $M(\beta)$ for any $\beta$, using the following lemma. 

\emph{Lemma}: If all eigenvalues of $M(\beta)$ have negative real parts for all values of $\beta$, then $\mathbf  u_0$ is linearly stable. \newline
\emph{Proof}: Given a time $t>0$ and an integer $n>0$, we define $t_i=\frac{t}{M}i$ for $i=0,\dots,n$. Thanks to the Mean Value Theorem, it is $\mathbf u_1(t_i)=\left(I+M_1\frac{t}{n}\right)\mathbf u(t_{i-1})$ for $i\ge 1$, where $M_i$ is the value of $M$ for a time in $(t_{i-1},t_i)$ and use has been made of Eq. \eqref{eq:87} to evaluate the time derivative. By iteration, $\mathbf u_1(t_i)=\left[\prod_{k=1}^i\left(I+M_k\frac{t}{n}\right)\right]\mathbf u_0$. Denoting by $\lVert\cdot\rVert$ any vector norm, we have $\lVert \mathbf u_1(t)\rVert = \lVert \left[\prod_{k=1}^n(I+M_k\frac{t}{n})\right]\mathbf u_0 \rVert \le \lVert (I+\tilde M\frac{t}{n})^n\mathbf u_0 \rVert$ where $\tilde M$ is such that $\lVert (I+\tilde M\frac{t}{n})\mathbf u_0 \rVert=\max_{k}\lVert\left(I+M_k\frac{t}{n}\right)\mathbf u_0 \rVert$. Taking $n\to \infty$, $\lVert \mathbf u_1(t)\rVert \le \lVert e^{\tilde M t}\mathbf u_0\rVert $ which tends to zero as $t\to \infty$, as all eigenvalues of $\tilde M$ have negative real parts. $\square
$

Using the lemma, we see that the only-cooperators solution is stable above line $\epsilon_c(p)$ given by Eq. \eqref{eq:78}, and the only-defectors solution is stable below line $\epsilon_d(p)$ given approximately by Eq. \eqref{eq:82}. This implies that the instability of the one-type solutions is due to the presence of coexistence solutions which become stable. Numerical evaluation of the time evolution of system \eqref{eq:37}-\eqref{eq:41} confirms the theoretical analysis.

\section{Discussion}
\label{sec:5}

As the root of the present work, we introduce a basic stochastic model of a spatially extended population of altrustic and non-altruistic agents, called cooperators and defectors. 
The population evolves by a birth-death process. In line with the considerations by Huang and colleages \cite{Huang:2015}, an agents' interaction with another agent influences the death rates only. 
Agents' interactions are altruistic acts. They lower the death rate of the recipient while increasing the death rate of the donor, relative to a baseline death rate $p$ for agents in isolation, $p$
being one of two parameters of the model. The benefit-cost ratio of the altruistic is encoded in the second parameter $\epsilon$.

Results are obained as (1) stochastic simulations of finite systems and (2) stationary solutions and their stability in approximate descriptions by rate equations. The pair approximation, neglecting all spatial correlations except those of nearest neighbors, yields our main result: For any benefit-cost ratio above $1$, the stable stationary solutions in dependence of baseline death rate $p$ display (i) a regime of co-existence of cooperators and defectors and (ii) a regime of a population of cooperators only. In the $(p,\epsilon)$ parameter plane, these regimes and the related transitions appear as a continuation of the known extinction transition for a spatial population without cooperative interaction (also known as contact process, asynchronous SIS model). The latter case corresponds to benefit-cost ratio of exactly 1 ($\epsilon=0$).

The phase diagram from pair approximation is fully qualitatively consistent with that from stochastic simulation with finite square lattices. 
Simulations of sufficiently large instances of $k$-regular random graphs yield an equivalent phase diagram (results not shown here); this holds also for preliminary simulation results on other graphs,
including scale-free \cite{baal99} and small-world networks \cite{wast98}. Thus we speculate that the observed type of $(p,\epsilon)$ phase diagram is generic, holding for most types of connected
sparse graphs. For dense graphs, however, we expect mean-field behavior without stable cooperation  seen in Sec.~\ref{sec:global_mean}.

Consider a spatially extended population subject to a decline in livability, which in reality may be a reduction of food resources or an increase of predators. In our model, this scenario is represented by increasing $p$ and comes with the following prediction. Initially without cooperators, the concentration of agents decreases until reaching a transition point with the onset of co-existence. In this
regime, the concentration of defectors further decreases; this decrease is overcompensated by the increase in cooperators. Thus in the co-existence regime, there is net population growth under increasing $p$
\cite{Sella:2000}. Further increase of $p$ first leads to a regime with a population containing cooperators only and then an extinction phase where zero population size is the only stable solution. 

From earlier studies, both theoretical \cite{Sella:2000} and experimental, increasing baseline death rate has been known to enhance cooperation. Perturbing populations of yeast cells by dilution shocks, S\'{a}nchez and Gore find populations with larger fractions of cooperative cells (providing digestive enzyme to the population) more likely to survive \cite{sago13}. Datta and co-authors observe cooperation promoted when a population expands the space it occupies, cooperators forming a wave of invaders \cite{dakocvdugo13}.
According to the rule by Ohtsuki and colleagues \cite{ohhalino06}, cooperation supersedes defection when the benefit-cost ratio is larger than the agent's number of neighbors $z$. While their theory assumes a population of constant size and each agent with a constant number $z$ of neighbors, we here see cooperation enhanced when the number of neighbors (occupied adjacent sites) is reduced dynamically due to a shrinking population density. 
When the population most ``needs'' it, i.e.\ at low density close to extinction, cooperation appears as a stable stationary solution for any benefit-cost ratio above 1. Future work may check if a rule relating benefit-cost ratio and neighborhood size characterizes the appearance of stable cooperation also in the present model with varying population size. For experimentally testing the present model's predictions, the unperturbed steady state of a population would have to be observed.

Giving up the spatial homogeneity of the baseline death rate $p$, we have investigated the scenario of a gradient between low $p$ (high livability) and high $p$ (low livability). The regimes encountered previously by tuning $p$ for the whole system are now found simultaneously at their corresponding spatial position. In particular, high concentration of cooperators is found next to the region uninhabited due to large death rate $p$. There is a region of co-existence where total population concentration increases with $p$ also spatially. Cooperation arises when and where needed to avoid extinction.

\section*{Acknowledgments}

Partial financial support has been received from the Agencia Estatal de
Investigacion (AEI, Spain) and Fondo Europeo de Desarrollo Regional
under Project PACSS RTI2018-093732-B-C21 (AEI/FEDER,UE) and the Spanish
State Research Agency, through the Maria de Maeztu Program for units of
Excellence in R\&D (MDM-2017-0711).
KK acknowledges funding from MINECO through the Ram{\'o}n y Cajal program
and through project SPASIMM, FIS2016-80067-P (AEI/FEDER, EU).

\bibliographystyle{unsrt}
\bibliography{coop_extinction}

\end{document}